\documentclass[submission,copyright,creativecommons]{eptcs}
\providecommand{\event}{Refine 2013} % Name of the event you are submitting to

\usepackage{mathbbol,graphicx,wrapfig,amsmath}
\usepackage{czt}
\usepackage{specQueues}
\usepackage{multicol}
\usepackage[font=footnotesize,caption=false]{subfig}

\zedsize{\small}
\renewcommand{\zedindent}{1em}

\newcommand{\refines}{\ensuremath{\sqsupseteq}}
\newcommand{\conj}{\ensuremath{\vdash?~}}

\newcommand{\melem}[1]{\texttt{#1}}

\title{Relaxing Behavioural Inheritance}
\author{Nuno Am\'alio
\institute{University of Luxembourg\\
6, rue Richard Coudenhove-Kalergi, L-1359, Luxembourg}
\email{nuno.amalio@uni.lu}
}
\def\titlerunning{Relaxing Behavioural Inheritance}
\def\authorrunning{N. Am\'alio}

\begin{document}

\maketitle

\begin{abstract}
Object-oriented (OO) inheritance allows the definition of families of classes in a hierarchical way. In behavioural inheritance, a strong version, it should be possible to substitute an object of a subclass for an object of its superclass without any observable effect on the system. Behavioural inheritance is related to formal refinement, but, as observed in the literature, the
  refinement constraints are too restrictive, ruling out many useful
  OO subclassings. This paper studies behavioural inheritance in
  the context of ZOO, an object-oriented style for Z. To overcome refinement's restrictions, this paper
  proposes relaxations to the behavioural inheritance refinement rules. The work is presented for Z, but the
  results are applicable to any OO language that
  supports design-by-contract.
\end{abstract}

\section{Introduction}

Object-oriented (OO) designs are structured around abstractions called \emph{classes}, which represent sets of objects with certain properties in common. OO \emph{inheritance}~\cite{Meyer:97} defines families of classes with a hierarchical structure, in which 
higher-level abstractions (superclasses) capture state and behavioural properties that all of its specialised abstractions (subclasses) have in common. Inheritance addresses \emph{reuse}, an important software engineering concern; in a hierarchy, subclasses reuse the behaviour of their superclasses, and add some specialised behaviour of their own.

Inheritance hierarchies have an \emph{is-a} semantics. A child abstraction (a subclass) is a kind of a parent abstraction. The child may have extra properties, but it has a strong conceptual link with the parent; an object of a descendant is at the same time also an object of the parent class (a parent class includes all objects that are its own direct instances plus those of its descendants). 

A consequence of the \emph{is-a} semantics is \emph{substitutability}: a subclass object can be used whenever a superclass object is expected. Substitutability is enforced in two different ways. Most OO systems enforce substitutability by checking \emph{interface conformity} using type-checking: the signatures of the subclass operations that specialise superclass operations must conform according to certain type rules. This guarantees that subclasses can be asked to do whatever their superclasses offer. However, a subclass may comply with the interface of its superclass, but it may go along and do something different. This problem is addressed by \emph{behavioural inheritance}~\cite{Subtyping:LW:1994}, a strong flavour of inheritance, which enforces substitutability by checking \emph{behavioural conformity} using proof: not only the interfaces must conform, the behaviour must conform also. This ensures that any subclass
object may replace an object of the superclass without any effect on
the superclass object's observable behaviour.

As observed in Liskov and Wing's seminal paper~\cite{Subtyping:LW:1994}, behavioural inheritance is related to \emph{data refinement}, which is also concerned with
substitutability. In~\cite{DataRef:HHS:1986}, Hoare et al define data
refinement as: ``One datatype (call it concrete) is said to refine
another data type [\ldots] (call it abstract), if in all
circumstances and for all purposes the concrete type can be validly
used in place of the abstract one.'' Inheritance relations should,
therefore, observe a refinement relationship between subclass
(concrete) and superclass (abstract). The difference is that whereas
in data refinement the refinement relation varies, in behavioural
inheritance this relation always follows the same pattern: a
function from subclass to superclass. Behavioural inheritance is, therefore, a specialisation of data refinement.

The drawback of data refinement models of inheritance is that they are over-restrictive. Liskov and Wing~\cite{Subtyping:LW:1994} mention: ``the requirement we impose is very strong and raises the concern that it might rule out many useful subtype relations''. 
This paper investigates behavioural inheritance and  proposes
relaxations to lift certain behavioural inheritance constraints and proof obligations. The investigation is in the
context of Z~\cite{UsingZ}, a formal modelling language based
on typed set theory and predicate calculus, with a mature
refinement theory~\cite{UsingZ,Refinement-Z-OZ}.  The work is part of
ZOO, the OO style for Z presented in~\cite{ZOO:2005,PhDThesis}, that is the semantic domain of the 
$UML+Z$ framework~\cite{PhDThesis,Amalio:2006aa} and the Visual Contract Language (VCL) for graphical modelling of software designs~\cite{VCL:VCB:2011,VCL:DBC:2010,VCL:TAOSD:2010}. 

 This paper's main contributions are three relaxations to facilitate use of rigorous behavioural inheritance in OO design, which are a result of a careful examination of mainstream OO inheritance in a Z data refinement setting using ZOO. The paper makes another contribution: it provides a way of specifying inheritance hierarchies in Z extending what is presented in~\cite{ZOO:2005} and that improves previous work.

\section{ZOO: A Z model of OO}

\begin{wrapfigure}[16]{l}{6.7cm}
\vspace{-4mm}
\includegraphics[scale=0.5]{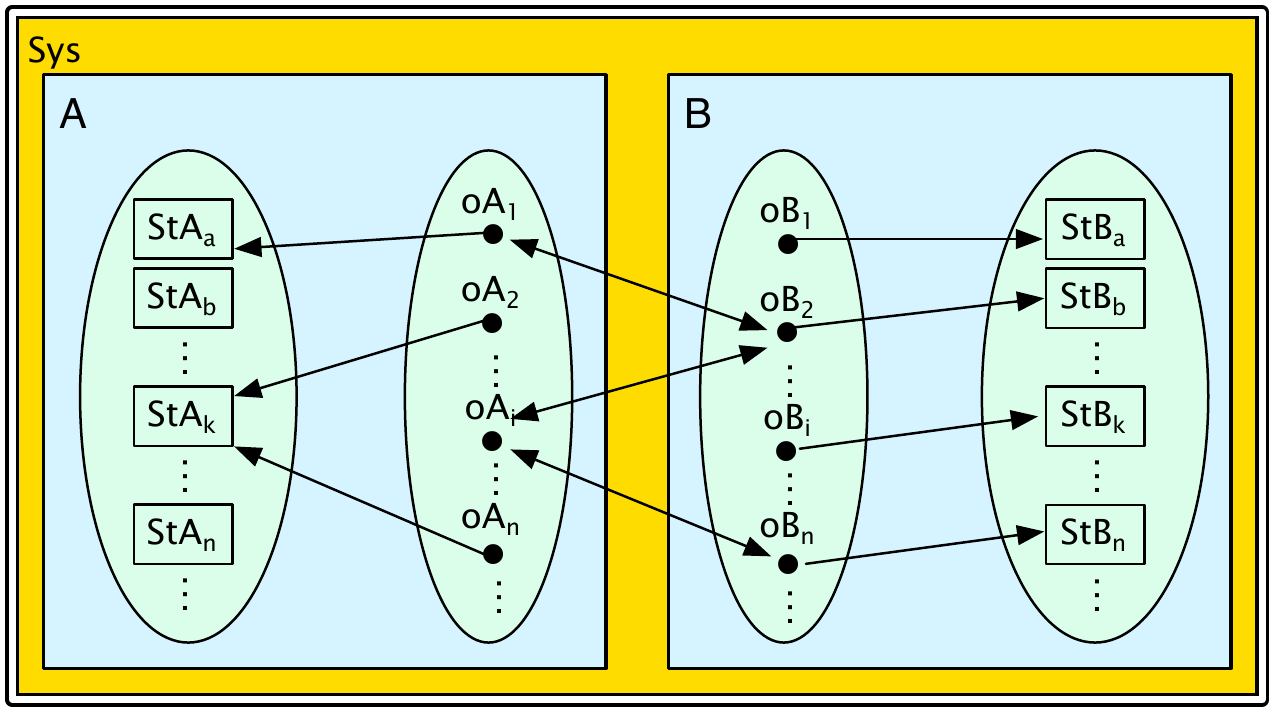}
\caption{A ZOO system ($Sys$) made of two classes and an association between them. Class \melem{A} comprises the set of all its object atoms ($OA$) and the set of all possible object states ($StA$); likewise for class \melem{B}. Association comprises a set of tuples relating objects of classes \melem{A} and \melem{B} (sets $OA$ and $OB$). A legend for the figure is given in Fig.~\ref{Fig:LegendZOO}}
\label{Fig:ZOOSys}
\end{wrapfigure}

ZOO~\cite{ZOO:2005,PhDThesis} is an approach to specify OO models in Z: a Z style of object-orientation. Its OO model is based on  Z abstract data types (ADTs) represented as Z schemas, constituting an OO model based on records~\cite{Cardelli:1988uq}. ZOO is an extension of Hall's OO Z style~\cite{Z-OO-Hall:1994,Z-OO-Hall:1990}.

\subsection{Overview}

ZOO represents objects as atoms. It considers that, like a set, a class has dual meaning. \emph{Class intension} defines a class in terms of the properties shared by its objects (for example, class \melem{Person} with properties \melem{name} and \melem{address}). \emph{Class extension} defines a class in terms of its currently existing objects (for example \melem{Person} is $\{MrSmith, MrAnderson, MsFitzgerald\}$).  This duality is expressed in terms of the representation of a class as a promoted Z ADT~\cite{UsingZ,Stepney-etal:2003:ZB} that is made up of an
inner (or local) type (the class intension), and an outer (or global) type containing the actual object instances and defining the interface to the environment (the class extension). ZOO represents OO associations between classes as binary relations between the sets of objects of each class. A system is a collection of classes and their associations. Figure~\ref{Fig:ZOOSys} illustrates ZOO's OO model with a OO system made of classes \melem{A} and \melem{B} with an association between them.

\subsection{Classes and Promotion}

Promotion~\cite{UsingZ,Stepney-etal:2003:ZB} is a technique to build composite structures in the Z schema calculus, 
so called because the inner type is
\emph{promoted} to a global state, without the need to redefine the
encapsulated ADT. Typically, the state of a promoted ADT is described as a
partial function $f : I \pfun S$,
where $I$ is a set of identities and $S$ a set of states. Promotion builds operations of an ADT $P$ in terms of operations of an encapsulated ADT
$S$ modularly (without changing $S$). In the context of a ZOO class, $I$ represents a set of
object identities of some class (the class extension), and $S$ represents the state space
of the class's objects (the class intention). This is depicted in Fig.~\ref{Fig:Promotion} and captured by ZOO's class 
generic: 
\begin{schema}{SCl}[OS, OST]
os : \power OS \\
oSt : OS \pfun OST
\where 
\dom~ oSt = os
\end{schema}
Here, the parameter $OS$ represents the set of possible objects of the class and $OST$ represents the set of possible states of the class (the class intension).

%In ZOO, we say that a promoted ADT
%comprises two views: intensional and extensional (by analogy with
%set theory). The inner type (the ADT $S$, above) corresponds to the
%intensional view; it defines the properties that are shared by all
%individual objects of the class. The outer type corresponds to the extensional
%view and defines a class in terms of its existing objects, where
%class operations (the interface to the outside world) are built by
%promoting operations of the inner type.

\subsection{Inheritance}

\begin{figure}[t]
\subfloat[A ZOO class as a Z promoted ADT]{\label{Fig:Promotion}
\includegraphics[scale=0.5]{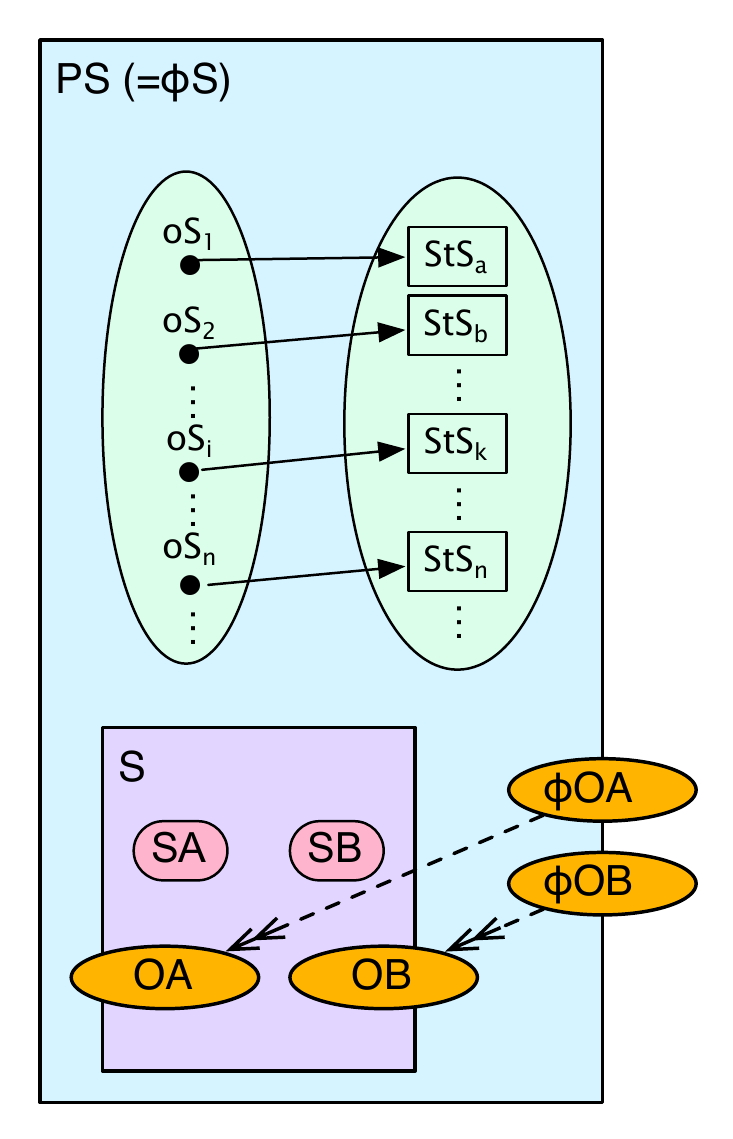}
}
\subfloat[The extensional viewpoint]{\label{Fig:ZOOInhClExt}
\includegraphics[scale=0.5]{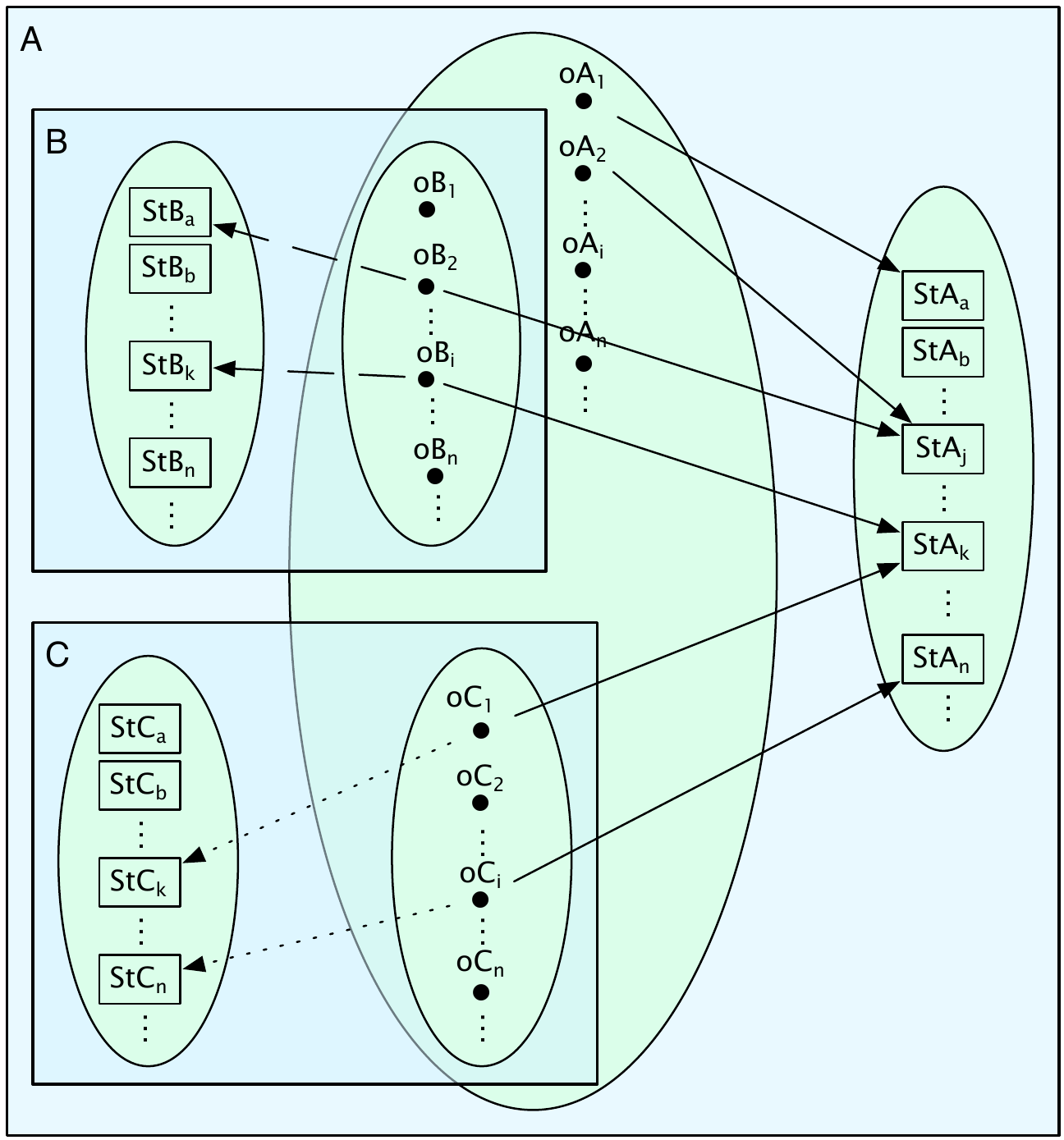}
}
\begin{tabular}[b]{l}
\subfloat[Legend]{\label{Fig:LegendZOO}
\includegraphics[scale=0.5]{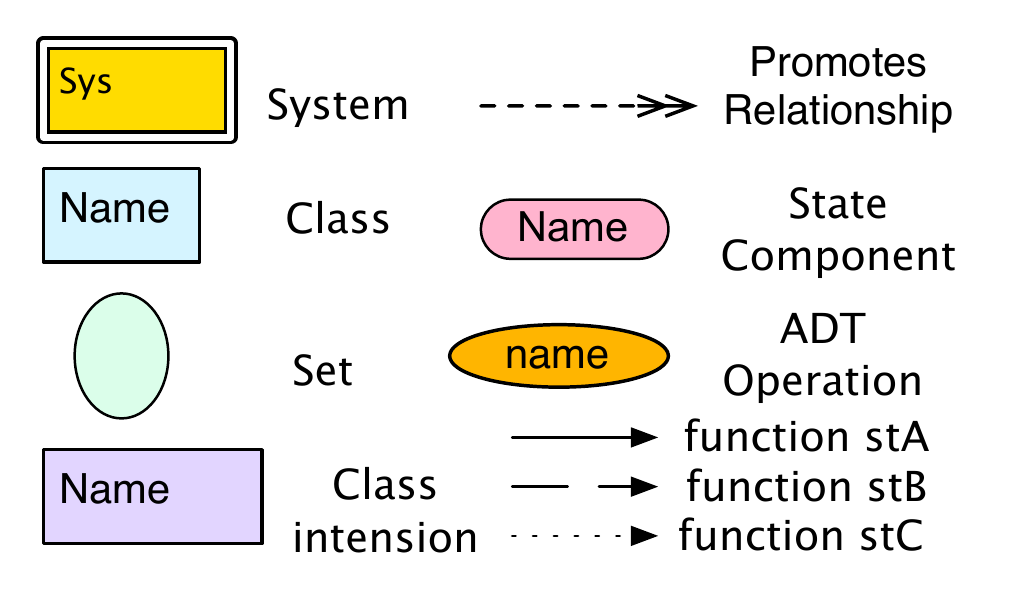}
}\\
\subfloat[The intensional viewpoint]{\label{Fig:ZOOInhClInt}
\includegraphics[scale=0.5]{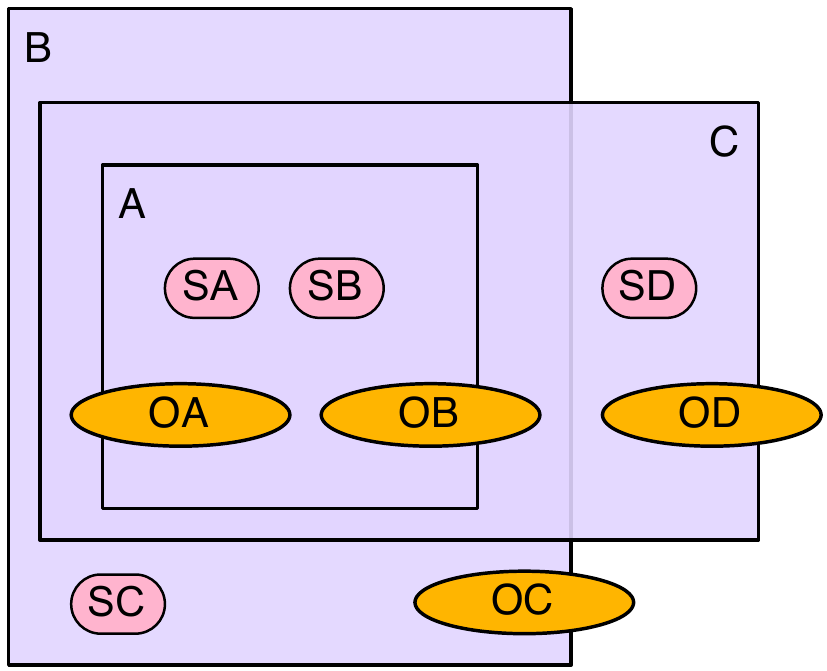}
}
\end{tabular}
\caption{ZOO's model of inheritance. In (a), a class \melem{PS} is represented as promoted Z ADT, separating its extensional and intensional  viewpoints. Class intension (d) and extension (b) in the context of OO inheritance are illustrated with a class \melem{A} and its subclasses \melem{B} and \melem{C}.
In extension (b), \melem{A} includes all its own objects ($OAs$) and all the objects of its subclasses ($OBs$ and $OCs$). Each class includes a set of all possible states of its objects ($StAs$, $StBs$, $StCs$); each class has its own mapping function. In intension (d), subclasses extend superclasses with further state (state components $SC$ and $SD$ in Fig. 2d) and operations ($OC$ and $OD$) using Z schema calculus conjunction. }
\label{Fig:ZOOInhCl}
\end{figure}

A OO model of inheritance needs to consider: (a) subclassing as subsetting, (b) subclass specialisation and (c) abstract classes and polymorphism. These concerns are address by ZOO's model of inheritance, depicted in Fig.\ref{Fig:ZOOInhCl}. 

\begin{itemize}
\item Extensionally, subclassing is subsetting. A subclass object is also an object of its superclasses; the sets of subclass objects are subsets of their superclasses. In Fig.~\ref{Fig:ZOOInhClExt}, class \melem{A} has subclasses \melem{B} and \melem{C}; set of \melem{A} objects comprises its own ($oA_i$), plus those of its subclasses ($oB_i$ and $oC_i$).
\item Subclasses specialise or extend the state and behaviour of their superclasses. This is emphasised in the intensional viewpoint of inheritance illustrated in Fig.~\ref{Fig:ZOOInhClInt}: classes \melem{B} and \melem{C} extend the state and behaviour components that they inherit from \melem{A}. In Fig.~\ref{Fig:ZOOInhClExt}, each subclass has its own state; the states of subclasses, however, extend the state of their superclasses. As each subclass object can been as an object of either superclass or subclass, there are mapping functions that map a subclass object to either superclass or subclass state (mapping functions in Fig.~\ref{Fig:ZOOInhClExt}). Constraints ensure that the states of subclass objects are kept consistent.
\item In ZOO, abstract classes do not have direct instances. If class \melem{A} in Fig.~\ref{Fig:ZOOInhCl} were abstract, then it would just consist of objects of classes \melem{B} and \melem{C}. Polymorphism refers to the ability of treating objects of abstract classes polymorphically: the actual behaviour of some objects depends on their direct classes. This is a matter of selecting the right behaviour given a superclass object (the next section shows how this is specified in ZOO).
\end{itemize}

\section{Specification of inheritance in ZOO}
\label{sec:spec-inh}
 
\begin{wrapfigure}[10]{l}{6cm}
\vspace{-0.6cm}
\begin{center}
%\begin{figure}
\includegraphics[scale=0.50]{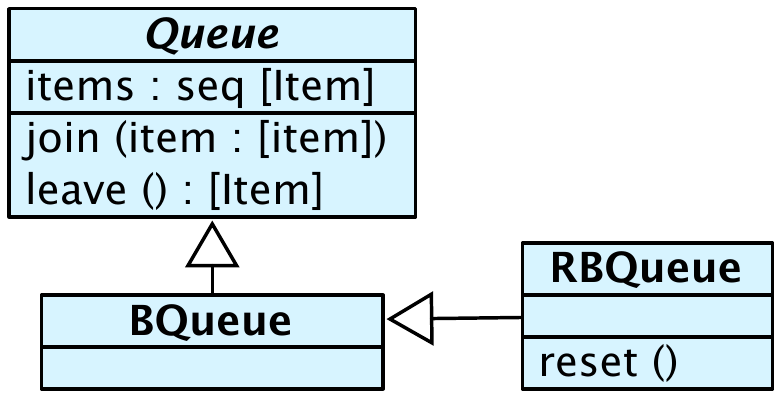}
\end{center}
\vspace{-0.5cm}
\caption{An inheritance hierarchy of queues formed by classes \melem{Queue}, \melem{BQueue} (bounded-queue), and
\melem{RBQueue} (resettable-bounded-queue).}
\label{fig:QBQRQ}
%\end{figure}
\end{wrapfigure}
ZOO's approach to inheritance is illustrated with a
class hierarchy of Queues (Fig.~\ref{fig:QBQRQ}). Class
\melem{Queue} stores a sequence of items (attribute
\melem{items}); it is an abstract class. It comprises two operations: \melem{join} adds an element to the queue, and \melem{leave} removes the element at front of the queue.  Class \melem{BQueue} (bounded queue) bounds the size of 
the queue. Class \melem{RQueue}
(resettable-bounded queue) introduces  the
extra operation \melem{reset}, to empty the
sequence of items. 

The following builds the ZOO model corresponding to Fig.~\ref{fig:QBQRQ} for each view of the ZOO style. The full model is given at \url{http://bit.ly/ftKZCp}. The behavioural inheritance examination conducted in this paper uses this hierarchy.

\subsection{Structural view}

The following defines the sets $CLASS$ (all class atoms ), $OBJ$ (all possible objects) and $abstractCl$ (all abstract classes), and the relation $subCl$ (subclass relation):

\noindent\begin{minipage}[t]{.35\linewidth}
\begin{zed}
CLASS ::= QueueCl | BQueueCl \\
\quad | RBQueueCl 
\also
[OBJ]
\end{zed}
\end{minipage}
\begin{minipage}[t]{.65\linewidth}
\begin{axdef}
abstractCl : \power CLASS\\
subCl : CLASS \rel CLASS\\
\where
abstractCl = \{QueueCl\}\\
subCl = \{BQueueCl \mapsto QueueCl,
   RBQueueCl \mapsto BQueueCl\}
\end{axdef}
\end{minipage}

In inheritance, the set of objects of a class includes its own
objects and those of its subclasses. Function $\oset$ gives
all possible objects of a class. Function $\osetx$ gives the direct set of objects of a class (excludes objects of subclasses). These two
functions are defined as:

\noindent\begin{minipage}[t]{.52\linewidth}
\begin{axdef}
\osetx : CLASS \fun \power_1 OBJ\\
\oset : CLASS \fun \power_1 OBJ
 \where
 \disjoint~ \osetx\\
  \forall cl : abstractCl @ \osetx~cl = \emptyset\\
\forall cl : CLASS @  \oset~cl = \osetx~cl \cup
\bigcup(\osetx~\limg~(subCl\plus)\inv \limg \{cl\}\rimg \rimg)
\end{axdef}
\end{minipage}
\begin{minipage}[t]{.48\linewidth}
\begin{zed}
\forall cl, cl' : CLASS | cl \mapsto cl' \in subCl @ \oset~cl
\subseteq \oset~ cl'
\end{zed}
\end{minipage}
Above to the left, the first axiom says that the sets of direct objects of each class are mutually
disjoint. The second says that abstract classes have an empty set of direct objects.
The third defines $\oset$ in terms of $\osetx$: 
the set of objects of a class includes its own objects and
those of its descendants. Above to the right, there is a useful law that can be extracted from the axioms to the left, which says that the set of all possible objects atoms of a subclass is a subset of its superclass counter-parts.

\subsection{Intensional view}

\subsubsection{Class \melem{Queue}}

This class comprises a sequence of items. Initially, the
sequence is empty. Operation \textsf{join} receives an item as
input and adds it to the back of the sequence. Operation
\textsf{leave} removes and outputs the item at the head of the sequence.
The intensional (or local) definition of \melem{Queue} is as follows:

\noindent\begin{minipage}[t]{.5\linewidth}
\begin{schema}{Queue}[Item]
items : \seq Item
\end{schema}
\end{minipage}
\begin{minipage}[t]{.5\linewidth}
\begin{schema}{QueueInit}[Item]
Queue~'[Item]
\where
items' = \langle \rangle
\end{schema}
\end{minipage}

\vspace{-5mm}

\noindent\begin{minipage}[t]{.5\linewidth}
\begin{schema}{\QueueEnter}[Item]
\Delta Queue[Item]\\
item? : Item
\where
items' = items \cat \langle item? \rangle
\end{schema}
\end{minipage}
\begin{minipage}[t]{.5\linewidth}
\begin{schema}{\QueueLeave}[Item]
\Delta Queue[Item]\\
item! : Item
\where
items \neq \langle \rangle
\also
item! = head~~ items
\land
items' = tail~~ items
\end{schema}
\end{minipage}

\subsubsection{Class \melem{BQueue}}

The intension of \textsf{BQueue} is defined by extending \melem{Queue}, its superclass. 
 The constant $maxQ$ defines the maximum number of items in the
queue. The invariant states that the sequence is bound by this constant.

\noindent\begin{minipage}[t]{.5\linewidth}
\begin{axdef}
maxQ : \nat_1
\end{axdef}
\vspace{-0.9cm}
\begin{schema}{BQueue}[Item]
 Queue [Item]
\where
\# items \leq maxQ
\end{schema}
\end{minipage}
\begin{minipage}[t]{.5\linewidth}
\begin{schema}{BQueueInit}[Item]
BQueue~'[Item]  \\
QueueInit[Item]
\end{schema}
\end{minipage}
\vspace{-1.3cm}
\noindent\begin{minipage}[t]{.5\linewidth}
\begin{schema}{\BQueueEnter}[Item]
\Delta BQueue [Item]\\
\QueueEnter
\end{schema}
\end{minipage}
\begin{minipage}[t]{.5\linewidth}
\begin{schema}{\BQueueLeave}[Item]
\Delta BQueue [Item]\\
\QueueLeave
\end{schema}
\end{minipage}

\subsubsection{Class \melem{RBQueue}}

Class \textsf{RBQueue} is defined similarly by extending \melem{BQueue}. \textsf{RQueue}'s
extra operation, \melem{Reset}, resets the
sequence of items to the empty sequence:

\vspace{-5mm}

\noindent\begin{minipage}[t]{.5\linewidth}
\begin{schema}{RBQueue}[Item]
 BQueue [Item]
\end{schema}
\end{minipage}
\begin{minipage}[t]{.5\linewidth}
\begin{schema}{RBQueueInit}[Item]
RBQueue~'[Item]; BQueueInit[Item]
\end{schema}
\end{minipage}

\vspace{-0.2cm}

\noindent\begin{minipage}[t]{.33\linewidth}
\begin{schema}{\RBQueueEnter}[Item]
\Delta RBQueue [Item]\\
 \BQueueEnter
\end{schema}
\end{minipage}
\begin{minipage}[t]{.33\linewidth}
\begin{schema}{\RBQueueLeave}[Item]
\Delta RBQueue [Item]\\
\BQueueLeave
\end{schema}
\end{minipage}
\begin{minipage}[t]{.33\linewidth}
\begin{schema}{\RBQueueReset}[Item]
\Delta RBQueue [Item]\\
\where
 items' = \langle \rangle
\end{schema}
\end{minipage}

\vspace{-3mm}
\subsection{Extensional View}
Class extensions of all classes (abstract and non-abstract) are defined like normal classes (see ~\cite{ZOO:2005}): by
instantiating the $SCl$ Z generic. State extensions of \melem{Queue}, \melem{BQueue}, and
\melem{RBQueue} are:
\begin{zed}
SQueue[Item] == SCl[\oset~ QueueCl, Queue[Item]][sQueue /os, stQueue /oSt]
\also
SBQueue[Item] == SCl[\oset~ BQueueCl, BQueue[Item]][sBQueue /os, stBQueue
  /oSt]
  \also
SRBQueue[Item] == SCl[\oset~ RBQueueCl, RBQueue[Item]][sRBQueue /os, stRBQueue
  /oSt]
\end{zed}

For each subclassing, there is a schema expressing the required constraints, namely:
(a) the set of existing objects of a subclass is a subset of
its superclass, and (b) the mapping functions of both classes must
be consistent. The subclassing schema for  
\melem{Queue}/\melem{BQueue} is:
\begin{schema}{SBQueueIsQueue}[Item]
SQueue[Item]; SBQueue[Item]
\where
sBQueue \subseteq sQueue \\
\forall oBQueue : sBQueue @ (\lambda~BQueue[Item] @ \theta Queue)(stBQueue~oBQueue) =
stQueue~oBQueue
\end{schema}
Here, the second conjunct of the predicate says that the inherited state of a \melem{BQueue} object must be the same no matter the object is seen as 
\melem{BQueue} or \melem{Queue}.

Operations of non-abstract classes are formed using promotion like those of normal classes (see ~\cite{ZOO:2005}).
The update operations of
\melem{BQueue}, defined from the promotion frame $\PhiSUpdBQueue$ defined below,  are: \label{Z:SQ:SRQ}
\begin{zed}
 SBQueueJoin[Item] == \exists \Delta BQueue[Item] @
\PhiSUpdBQueue[Item] \land \BQueueEnter[Item]
\also
 SBQueueLeave[Item]
== \exists \Delta BQueue[Item] @ \PhiSUpdBQueue[Item] \land
\BQueueLeave[Item]
\end{zed}

Promotion frames of subclass operations need
to take the \emph{subsetting} constraint into account. There
is an intermediate frame to specify the action in the
superclass, there are intermediate frames in the subclasses that
extend the superclass frame. The intermediate frames for
\melem{Queue} and \melem{BQUeue} are:

\vspace{-1mm}
\noindent\begin{minipage}[t]{.47\linewidth}
\!\!\!\!\begin{schema}{\PhiSUpdQueueZ}[Item]
\!\!\!\!\Delta SQueue[Item]\\
\!\!\!\!\Delta Queue[Item]\\
\!\!\!\! oQueue? : \oset QueueCl
\where
\!\!\!\! sQueue' = sQueue\\
\!\!\!\! stQueue' = stQueue 
  \oplus \{oQueue? \mapsto \theta Queue~'\}
\end{schema}
\end{minipage}
\begin{minipage}[t]{.53\linewidth}
\begin{schema}{\PhiSUpdBQueueZ}[Item]
\!\!\!\!\PhiSUpdQueueZ[Item][oBQueue?/oQueue?]\\
\!\!\!\!\Delta SBQueueIsQueue[Item]\\
\!\!\!\!\Delta BQueue[Item]\\
\!\!\!\! oBQueue? : \oset~ BQueueCl
\where
\!\!\!\! sBQueue' = sBQueue\\
\!\!\!\! stBQueue' = stBQueue 
  \oplus \{oBQueue? \mapsto \theta BQueue~'\}
\end{schema}
\end{minipage}
Here, the predicate of both \emph{update} frames says that the set of existing objects remains the same, and that the state function
is updated (using function overriding) for the updated object with the updated state.

The subclass final promotion frame extends the intermediate
frame with the required precondition:
\begin{schema}{\PhiSUpdBQueue}[Item]
\PhiSUpdBQueueZ[Item]
\where
 oBQueue? \in sBQueue \cap \osetx~BQueueCl\\
 \theta BQueue = stBQueue~oBQueue?
\end{schema}
Note that the promotion frames of the subclass ensure satisfaction of the subsetting
constraint: whenever an object is added to the subclass it is also
added to the superclass. 

Polymorphic operations are specified as choice of behaviours (a disjunction). They are built in a bottom-up fashion. Polymorphic operation \melem{BQueue.join} offers a choice between \melem{BQueue} and \melem{RQueue}:

\vspace{-0.5cm}
\begin{zed}
\SBQueueEnterP[Item] == \SBQueueEnter[Item] \lor \SRQueueEnter[Item]
\end{zed}

The operation \melem{join} on \melem{Queue} offers the polymorphic operation of \melem{BQueue}:

\vspace{-0.5cm}
\begin{zed}
\SQueueEnter[Item] == \SBQueueEnterP[Item]
\end{zed}

\subsection{Global View}

The system schema includes all class extensions and subclassing schemas:

\vspace{-0.5cm}
\begin{schema}{System}[Item]
SQueue[Item]; SBQueue[Item]; SRBQueue[Item]
\where
\SBQueueIsQueue[Item]
\land
\SRBQueueIsBQueue[Item]
\end{schema}

\section{Behavioural inheritance and Z Data Refinement}
\label{sec:beh-inh}

As discussed above, the correctness of inheritance hierarchies with respect to substitutability (behavioural inheritance) is checked using data refinement methods. This enables the application of the theory of data refinement, which is mature and well-developed, to the setting of OO design.  

\begin{figure}
\subfloat[Data refinement simulation]{\label{Fig:RefSimulation}
\includegraphics[scale=0.6]{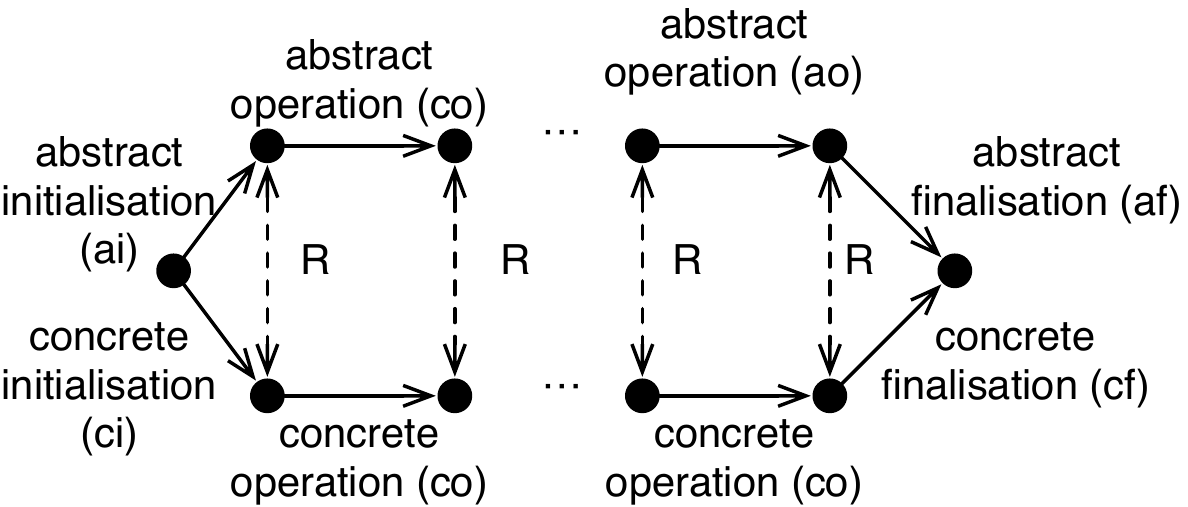}
}
\subfloat[Simulation in the context of promotion refinement]{\label{Fig:PromotionSimulation}
\hspace{2cm}\includegraphics[scale=0.6]{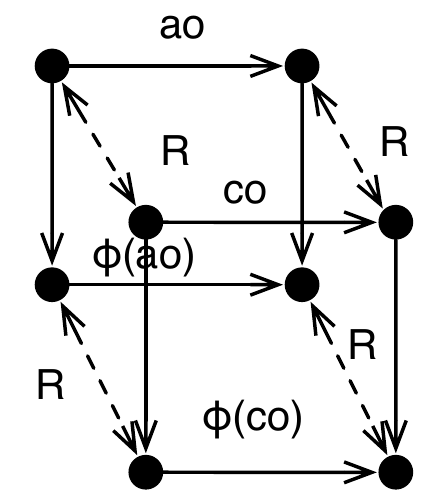}\hspace{2cm}
}
\caption{Simulation in the context of data refinement.}
\end{figure}

In Z, the correctness of a refinement is demonstrated using the concept of a \emph{simulation} (Fig.~\ref{Fig:RefSimulation}) with the aim of comparing ADTs inductively on a step by step basis~\cite{DataRef:HHS:1986}. 
This means that for each operation in the abstract type, there must be a corresponding operation in the concrete type. The correctness of the refinement involves proving
certain conjectures, known as simulation rules. The setting for
refinement in Z is as follows: (a) find a simulation relation relating concrete and abstract
data types, (b) demonstrate the correctness of the refinement by proving
the required conjectures. The conjectures vary with the type of
relation and the setting of refinement. A forwards (or downwards) simulation establishes a
map from the concrete to the abstract type; and a
backwards (or upwards) simulation is the other way
round~\cite{DataRef:HHS:1986}. There are two settings for refinement
in Z~\cite{Refinement-Z-OZ}: \emph{non-blocking} (contractual)
refinement interprets an operation as a contract and so outside the
precondition anything may happen, whilst \emph{blocking} (behavioural)
refinement says that outside the precondition an operation is blocked.

ZOO's inheritance model allows Z data refinement to be applied to the OO setting:
class refinement is simply the data refinement of the class's inner
and outer ADTs. In Z, this is well studied, and known as
\emph{promotion refinement}~\cite{Refinement-Z-OZ,UsingZ,Lupton:90} (Fig.~\ref{Fig:PromotionSimulation}).
 One result of promotion refinement is particularly useful: under certain circumstances,
promotion is compositional with respect to
refinement~\cite{Lupton:90}. That is, a promoted ADT refines another
if there is a refinement between the types being promoted. Formally,
suppose promoted ADTs $PC$ and $PA$ promote, respectively, $C$ and
$A$; then to prove that $PC$ refines $PA$ it is sufficient to show
that $C$ refines $A$. This applies when the promotion is \emph{free}
(discussed  below).

Next sections study behavioural inheritance in the context of the inner type (or
class intension). Behavioural inheritance conformance is checked by proving the
correctness of some refinement. 

%The following
%briefly reviews Z data refinement (section~\ref{sub:sec:Z-Ref}),
%explains the behavioural inheritance refinement relation (section~\ref{sub:sec:ref-relation}),
%derives refinement rules for behavioural inheritance
%(section~\ref{sub:sec:ref-rules-beh-inh}), discusses the special case
%of subclass extra operations
%(section~\ref{sub:sec:extra-operations}) and compares behavioural inheritance refinement rules derived here
%with the original rules of Liskov and Wing~\cite{Subtyping:LW:1994}
%(section~\ref{sub:sec:comparison}).

\subsection{A Refinement Relation for Behavioural Inheritance}
\label{sub:sec:ref-relation}

To define a particular subclassing at the local level (inner type or intension), the subclass schema extends its superclass  using Z schema
conjunction. Formally, for a class \textsf{A} (abstract) and its
subclass \textsf{C} (concrete), the state of \textsf{C} is defined in the intensional view by the following Z schema calculus formula, where $X$ represents the extra state of $C$.  In this setting, we can describe the relation between a subclass and its superclass as the function $f$:

\vspace{-1mm}
\noindent\begin{minipage}[b]{.5\linewidth}
\begin{zed}
C == A \land X
\end{zed}
\end{minipage}
\begin{minipage}[b]{.5\linewidth}
\begin{zed}
f = \lambda C @ \theta A
\end{zed}
\end{minipage}
This total function %takes an instance of $C$ (subclass state) and
% returns an instance of $A$ (superclass state). That is, it
projects the subclass state in terms of the superclass state, removing
the state added in the subclass (referred to as subclass-extra state).

\subsection{Refinement rules for behavioural inheritance}
\label{sub:sec:ref-rules-beh-inh}

Refinement rules for the above refinement function are derived in~\cite{PhDThesis}. For backward and forward simulation, the rules
reduce to a single set (unlike the general case, where
the simulations have separate rule sets). However, as expected, some simulation rules differ for blocking and non-blocking refinements.

Let $A$ and $C$ be class intensions defined in Z such that $C$ extends $A$ (i.e. $C = A \land X$). Let $A$ and $C$ have initialisation schemas $AI$ and $CI$, operations $AO$ and $CO$, and finalisation schemas $AF$ and $CF$\footnote{The finalisation condition describes a condition for the deletion of objects; e.g. a bank account may be deleted provided its balance is $0$.}.  For  non-blocking (NB) refinement, $C$ conforms to the behaviour of $A$, ($C \refines A$), if and only if:

\begin{enumerate}
\item $\conj \forall C~' @ CI \implies AI$ \hfill(Initialisation)
\item $\conj \forall C; i? : I @ \pre AO \implies \pre CO$ \hfill (Applicability)
\item $\conj \forall C~'; C; i? : I; o! : O
@ \pre AO \land CO \implies AO$ \hfill (NB Correctness)
\item $\conj \forall C @ CF \implies AF$ \hfill (Finalisation)
\end{enumerate}
In a OO setting, $A$ above corresponds to a superclass and $C$ to a subclass. The first rule allows subclass initialisations to be strengthened. The second rule allows the precondition of a subclass operation ($CO$) to
be weakened. The third rule says that the subclass operation must
conform to the behaviour of the superclass operation ($AO$) whenever
the superclass operation is applicable; this means
that the postcondition may be strengthened. The last rule allows the
finalisation to be strengthened; if the finalisation is total (the
ADTs do not have a finalisation condition) the fourth rule reduces to
$true$.

In the blocking setting, the correctness rule is strengthened to
require the precondition to remain the same:

\vspace{0.2cm}

\begin{list}{3a.}
\item $\conj \forall C~'; C; i? : I; o! : O @ CO \implies AO$ \hfill (B Correctness)
\end{list}

\vspace{0.2cm}

These rules dictate the proofs necessary for subclass
initialisation, finalisation and operation specialisations (that is,
operations that exist in the superclass), but not subclass-extra operations. 

\subsection{Extra operations}
\label{sub:sec:extra-operations}

Data refinement simulation requires that for each
valid execution in the concrete type there is a corresponding
execution in the abstract. Each execution step in the concrete type
must be simulated by the abstract. Thus, when a new operation is
added to a subclass, the refinement proof needs to show that the new
operation (concrete) simulates something in the superclass (abstract).

A common approach to ensure that refinement holds is to
routinely include in the abstract model an operation that does
nothing (called a \emph{stuttering} step or a \melem{skip}
operation), and then prove that the new concrete operation refines
\melem{skip}. The intuition is simple. Consider an ADT as a machine operated by buttons; the user presses a button
to execute an operation. In the abstract type, the
\melem{skip} operation button  exists but does nothing; in the
concrete type, the button executes the new operation.

The rules for checking subclass-extra operations are obtained from
the rules above by replacing $AO$ with \melem{skip} ($\Xi A$ in Z).This imposes a constraint
that the state is not changed by the operation.

\subsection{Liskov and Wing~\cite{Subtyping:LW:1994} revisited}
\label{sub:sec:comparison}

The rules above are consistent with those of Liskov
and Wing~\cite{Subtyping:LW:1994}. They allow any function between subtype and supertype as subtyping is not necessarily inheritance; here there is only one function to reflect the specific inheritance setting. Their rules correspond to the blocking rules presented above without initialisation and finalisation. There are
similar rules for extra operations, which are either a
combination of those in the superclass or change 
subclass-extra state  only (like \textsf{skip}).

\section{The refinement straight-jacket}
\label{sec:refinement-constraints}

The refinement rules of the previous section are over-restrictive. Common inheritance relations of OO design, such as the queues example of section~\ref{sec:spec-inh}, cannot be deemed behavioural inheritance conformant.

% using the refinement rules of the previous section.

% using the Queues example of section~\ref{sec:spec-inh} and the proof rules derived in section~\ref{sec:beh-inh}.

%Section~\ref{sec:beh-inh} revisited the theory of refinement to derive a set of proof rules for checking the conformance of behavioural inheritance in ZOO models. We now see  This section explains the problems encountered (sections
%\ref{sub-sec:s-extra-ops} and \ref{sub-sec:s-constraints}), and considers some existing ways in which
%these restrictions can be addressed.
%The next section discusses how the refinement rules can be relaxed.

\subsection{Subclass-extra constraints}
\label{sub-sec:s-constraints}

In the inheritance hierarchy of Fig.~\ref{fig:QBQRQ}, class \melem{BQueue} does not refine \melem{Queue}. The applicability conjecture for the operation
\melem{join} does not hold. The precondition of
\melem{Queue.join} is, $true$, whilst that of
\melem{BQueue.join} is $\# items < maxQ$. The former does not imply the latter and so applicability fails.
In refinement, the concrete type may weaken the precondition; here,
the subclass precondition is stronger.

That the refinement proof fails is reasonable. An operation is a
contract to the outside world. Abstract operation \melem{join}
contracts to do its job under any circumstance.
The concrete operation, however, introduces a pre-condition. This violates substitutability, because the
behaviour is observably different when the concrete type is used in
place of the abstract one.  Imagine a braking
system of a car, where the abstract type says ``upon brake slow down
in any circumstance'' (precondition $true$), and the concrete type
says, ``upon brake slow down only when speed is less than 160 Km
per hour'' (precondition $speed<160$); the concrete type is
obviously not a valid substitute of its abstract counter-part.

\subsection{Subclass-extra operations changing inherited state}
\label{sub-sec:s-extra-ops}

The inheritance refinement proof for \melem{RBQueue} also fails. The
\melem{reset} operation does not refine \melem{skip}: the correctness
conjecture is not provable because \melem{reset} violates the
 $\Xi Queue$ constraint by updating the inherited attribute $items$.

%This illustrates a conflict between the restrictions of
%refinement, and a common practice in OO  abstraction specialisation and
%incremental modification.

\subsection{Working round the restrictions of refinement}

There are two ways to address the problems imposed by the refinement
restrictions: (a) we can refactor the OO models to conform to the
refinement requirements, or (b) we can relax the formal restrictions.

Refactoring, a common software engineering practice, seeks to change a
model whilst preserving its meaning.
In this particular example, however, the refactoring would involve merging
the behaviours of the three classes into a single class.
The  \textsf{reset} operation changes
abstract state, so it needs moving in to the
superclass. \textsf{BQueue}'s specialised behaviour
would also need to be moved to the superclass.
This is a valid refactoring, albeit cumbersome: we lose the flexibility and modularity that inheritance provides.

\vspace{-2mm}
\section{Relaxations to the refinement constraints}
\label{sec:ref-relaxations}
\vspace{-2mm}

\begin{figure}\hspace{-2mm}
\subfloat[Classical Refinement]{\label{Fig:ClassicalRef}
\includegraphics[scale=0.5]{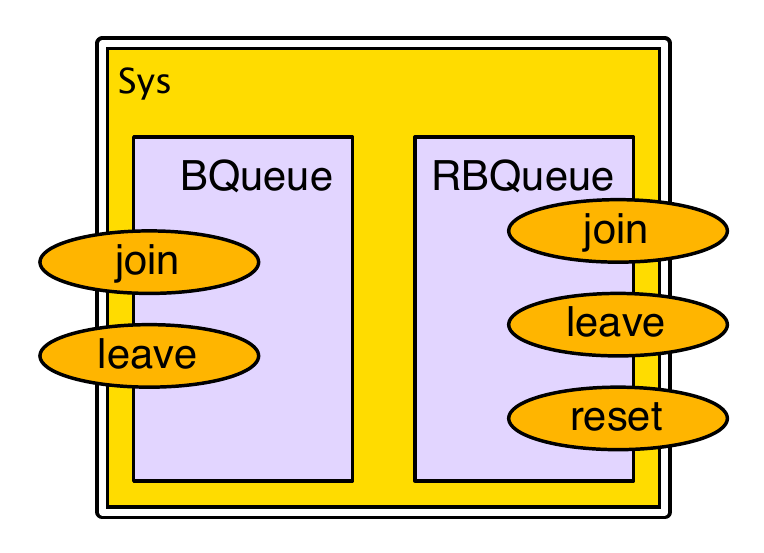}
}
\subfloat[Setting of OO Refinement used in this paper]{\label{Fig:OORefinement}
\includegraphics[scale=0.5]{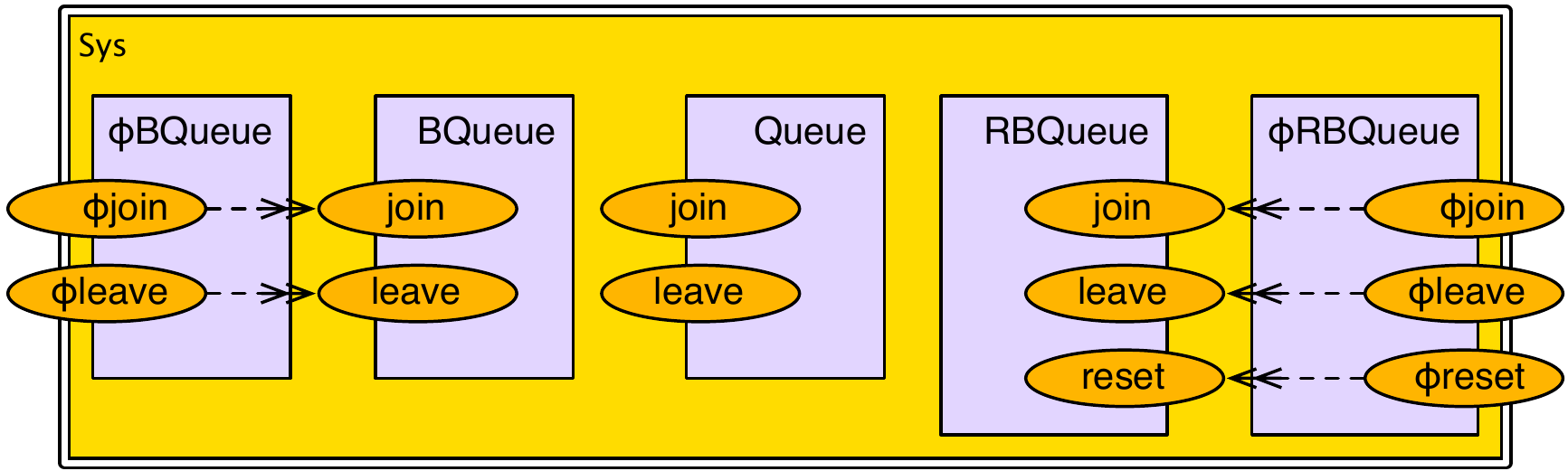}
}
\subfloat{
\includegraphics[scale=0.5]{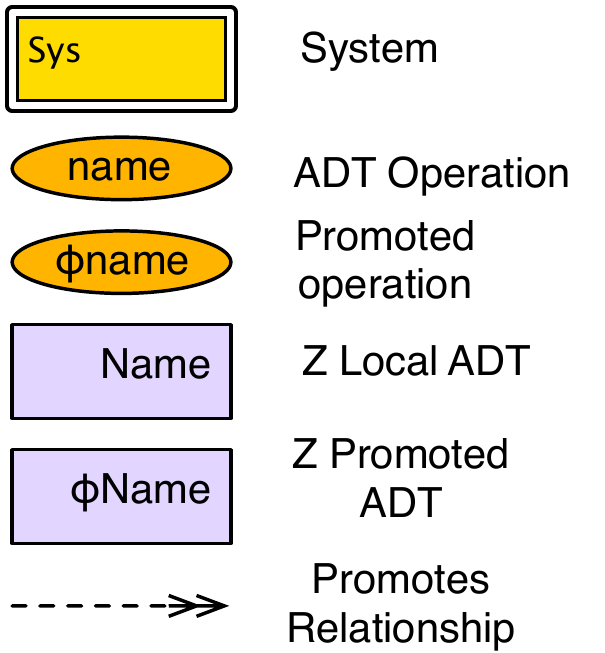}
}
\caption{Classical setting of data refinement vs OO-based data refinement used here illustrated with the queues example. Classical setting (a) assumes that operations of ADTs are exposed to the environment. In this paper's OO setting (b), it is the promoted operations of a promoted ADT (a class) that are visible to the environment; not all operations are necessarily promoted and, hence, not all of them are visible to the environment.}
\label{Fig:ClassicalvsOORefinement}
\end{figure}

The relaxations presented here exploit the specificities of the OO-based data refinement setting (Fig.~\ref{Fig:ClassicalvsOORefinement}). The general theory of data refinement (the classical setting) assumes that ADTs are exposed to the system environment (Fig.~\ref{Fig:ClassicalRef}): 
operations of ADTs (buttons in our
metaphor) are used by the environment to
interact with the system. However, ADTs are often concealed from the
environment and only used internally and this applies to the OO setting based on Z promotion used here (Fig.~\ref{Fig:OORefinement}): the
inner ADT (class intension) is concealed; some operations of inner ADTs may not be promoted and are, hence, also concealed. The following relaxations to behavioural inheritance exploit this.

\subsection{Relaxing with virtual superclass operations}

%\begin{wrapfigure}[18]{l}{4.5cm}
%\vspace{-3mm}
\begin{figure}
\subfloat[Extra sub-class operations]{\label{Fig:RefExtraOperations}
	\includegraphics[scale=0.5]{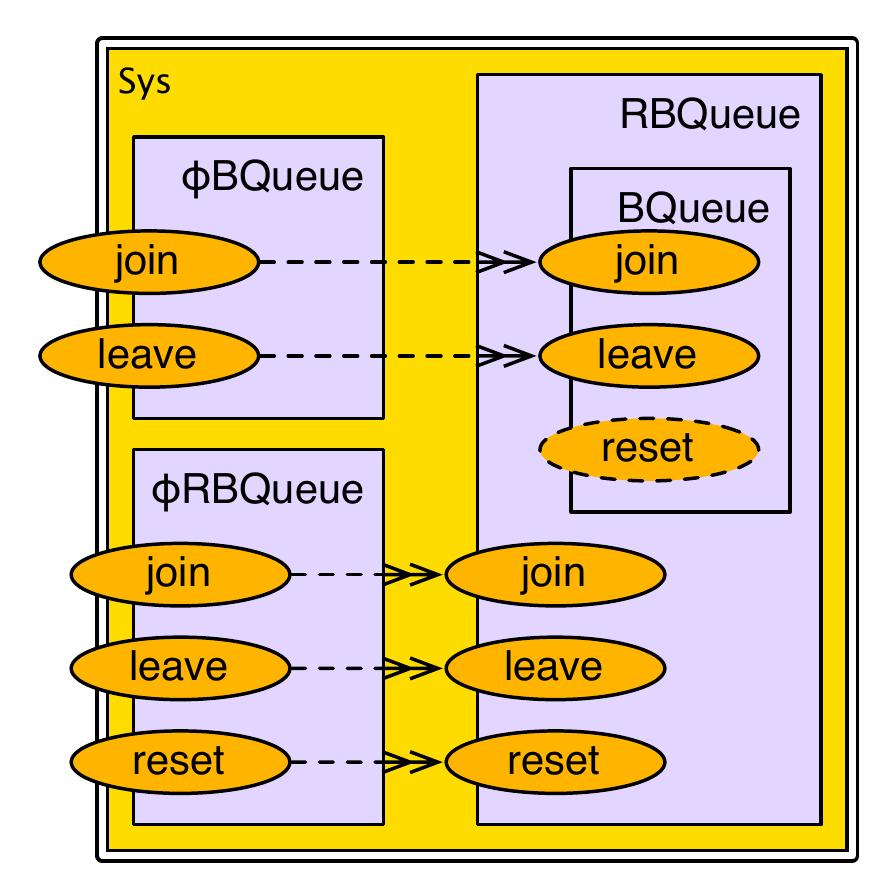}
}
\subfloat[OO abstract classes]{\label{Fig:RefAbstract}
	\includegraphics[scale=0.5]{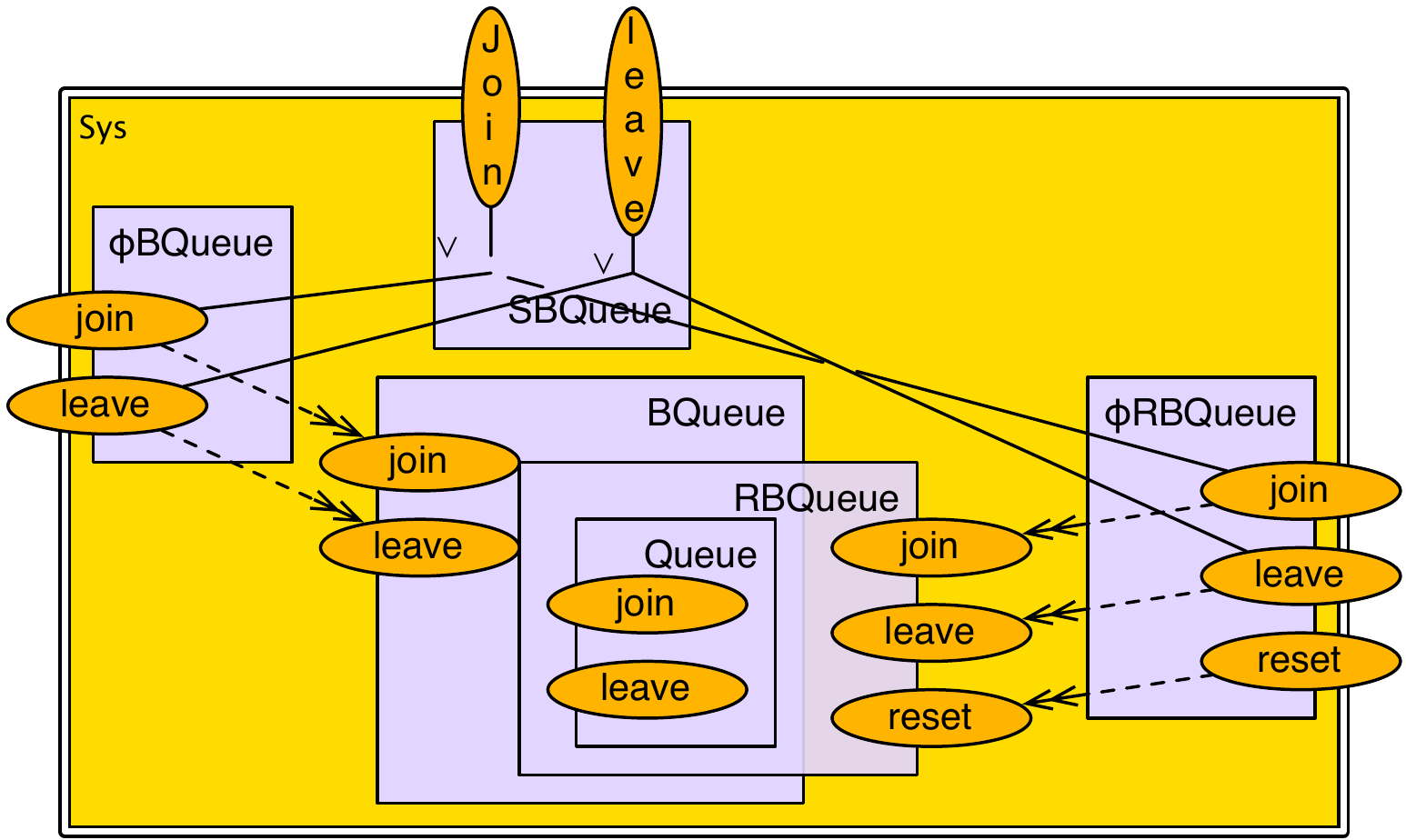}
}
\subfloat{
	\includegraphics[scale=0.5]{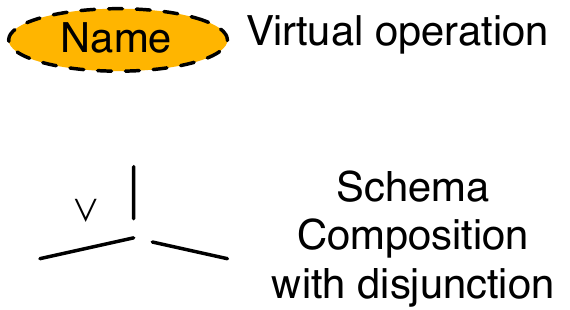}
}
\caption{Behavioural inheritance relaxations. In relaxation for extra sub-class operations (a), extra operation is simulated by an internal virtual operation that is never promoted and not called from the environment. In relaxation based on OO abstract classes (b), the inner ADT of the abstract superclass is never called from the environment so we can lift the applicability constraint. The operations of the abstract superclass that are made available to the environment are polymorphic (a disjunction).}
\end{figure}

As discussed in section~\ref{sub:sec:extra-operations}, the \melem{skip} approach preserves substitutability
to the environment: if pressed on the abstract type it does nothing;
 if pressed on the concrete type it does something, but respects
the behaviour set by the abstract type. Because of restrictions of
simulation in data refinement, we cannot just eliminate \melem{skip}. What we need is to
find a more \emph{liberal} replacement. This can be found by
exploiting the following: in the setting of OO inheritance a \melem{skip}-like button will never be pressed because it is not made
available to the environment (Fig.~\ref{Fig:RefExtraOperations}). For example, when \melem{RBQueue} is
used when a \textsf{BQueue} is expected, all is needed are 
operations \melem{join} and \melem{leave}: the simulating
substitute of operation \melem{reset} is never called from the
environment (Fig.~\ref{Fig:RefExtraOperations}).

The relaxation requires a superclass \emph{virtual} operation to simulate the operation in the subclass. In
general, given a subclass-extra operation $co$ there is a
virtual superclass operation $ao$ that simulates its behaviour. The refinement function enables the calculation of this virtual operation.  Briefly, given a subclass operation (concrete):
\begin{zed}
co = \{CO @ \theta C \mapsto \theta C~'\}
\end{zed}
\noindent the required superclass virtual operation (abstract) is given by the
formula:
\begin{zed}
ao = f\inv \comp co \comp f
\end{zed}

In~\cite{PhDThesis}, it is proved that any concrete operation ($co$) refines the calculated abstract operation ($ao$). This means
that subclass extra operations can be added freely: for any concrete operation $co$ there is always an abstract operation $ao$ that simulates it!

Using this relaxation, \melem{RBQueue} becomes behavioural
inheritance conformant with \melem{BQueue}. \melem{BQueue}'s extension provides only two operations to the environment,
\melem{join} and \melem{leave}.  Internally, \melem{RBQueue}
provides those operations and \melem{reset}; the calculated virtual
operation simulates \melem{reset} in \melem{BQueue}'s intension and is not made available to the environment (Fig.~\ref{Fig:RefExtraOperations}).

%We consider that \textsf{RBQueueN} is a \emph{strong} behavioural inheritance refinement of
%\textsf{BQueueN} because the relaxation to \textsf{skip} that guarantees this
%does not compromise the principles of data refinement.

\subsection{Relaxing by using OO abstract classes}

%\begin{wrapfigure}[20]{l}{7cm}
%\vspace{-5mm}
%\includegraphics[scale=0.5]{img/RefAbstract}
%\caption{}
%\label{Fig:RefAbstract}
%\end{wrapfigure}

A second relaxation exploits OO \emph{abstract}
classes\footnote{\emph{not} to be confused with a class that is abstract
  in the context of formal refinement!}. In OO, an abstract class has no
direct instances. Its operations definitions provide a basis
for reuse and polymorphism, but are never called from the environment; they are inherently \emph{virtual}. This is the basis of the relaxation.

The relaxation is: \emph{if a subclass inherits from an OO
  abstract class, the applicability proof obligation is lifted.}
The operation of an abstract class binds behaviour for its
descendants, so we need to prove correctness, but there is no need to prove applicability because OO-abstract superclasses
are never called from the environment.  In the button analogy, when the superclass is
OO-abstract, only its non-abstract subclasses can offer buttons to the environment;
the operations of the abstract class are inherently virtual (Fig.~\ref{Fig:RefAbstract}). This relaxation makes \melem{BQueue} behavioural-inheritance conformant with \melem{Queue}.

Care is needed when using OO abstract classes for relaxation: the precondition of the operation definition of an abstract class should not be relied upon to set an applicability behaviour for its descendants.

%\subsection{Queues example revisited}

%In the queues example of figure~\ref{fig:QBQRQ}, class \melem{Queue} is 
%OO-abstract. By using the abstract-class relaxation proposed here, this removes the need to prove applicability
%conjectures: Using the virtual superclass operations proposed here, \melem{RBQueue} becomes behavioural inheritance conformant with \melem{BQueue}. 

%The refinement checks performed so far, have been done at the local level of individual objects (class intension).
%The next section shows how these proved properties can be violated by global constraints.

\vspace{-4mm}
\section{The effect of global constraints}
\label{sec:outer-type}
\vspace{-2mm}

The previous section demonstrated the behavioural inheritance conformance of the Queues 
hierarchy (Fig.~\ref{fig:QBQRQ}) using the proposed relaxations.  The analysis was done in a local scope, where individual objects are isolated from other objects of the system. This section analyses behavioural inheritance under a more
global perspective to investigate the interference of global constraints.

\subsection{Promotion Refinement Revisited}

As mentioned in section~\ref{sec:beh-inh}, promotion is compositional with respect to refinement provided it is
\emph{free}. In the OO context exploited here, this means
that, provided the promotions are free, checking behavioural
conformance at the local level is sufficient to conclude conformance for the whole system.

A Z promotion is free if the inner type is not constrained from the
global space. The following analyses the \emph{freeness} constraint for the
queues example, before showing how the freeness
constraint can be relaxed so that the compositionality result is applicable to a
wider range of situations.

\vspace{-1mm}
\subsection{When inner behavioural conformance is not sufficient}
\vspace{-1mm}

Consider the
classes \melem{BQueue} and \melem{RBQueue} of figure~\ref{fig:QBQRQ}. Suppose,
we introduce a global constraint that affects the state of the
objects of \melem{RBQueue}, namely, the size of the sequence must be
strictly less than 2:
\begin{schema}{\SRBQueue}[Item]
 \classGen[\oset~ RBQueueCl, RBQueue][sRBQueue /os, stRBQueue
  /oSt]
\where \forall o : sRBQueue @ \#(stRBQueue~o).items < 2
\end{schema}

The new constraint violates behavioural inheritance, because, under certain circumstances, the behaviour of
\melem{RBQueue} objects diverge from 
\melem{BQueue} objects.  Suppose that we create objects $oQ$
of class \textsf{BQueue} and $oRQ$ of class \melem{RBQueue} (both
queues are empty).  If we execute operation \melem{join}  twice
on them, the observed behaviour is the same.  However, a
third call to \melem{join} on $oQ$ allows the object to be
added to the sequence, but fails on $oRQ$ because the call is outside the
precondition (there are already two items in the queue).  In the
non-blocking interpretation, any outcome is permitted, whilst in the
blocking interpretation, the operation blocks.  Substitutability is violated: $oRQ$ cannot be used in place of $oQ$.

%Although this illustration is contrived, it illustrates our point.
%We can imagine the same happening with other kinds of global
%constraints (such as association multiplicity constraints, or
%constraints relating the inner states of different classes).

% . After all, the global constraint
% can be described in terms of the inner type, in which case the two
% classes would not be behavioural conformant because applicability
% would not be provable\footnote{The only way to make them behavioural
% conformant would be to make \textsf{Queue} abstract and then appeal
% to our abstract-class relaxation}.

%The next section relaxes the freeness rule, so that we can rely on the compositionality property of promotion refinement.

\vspace{-1mm}
\subsection{Relaxing the freeness rule}
\vspace{-1mm}

The relaxation to the freeness rule takes the form of a design
guideline: \emph{global constraints should be expressed in terms of
  superclasses, otherwise they may interfere with behavioural
  conformance proofs.} So we have relaxed from ``promotion refinement
is compositional provided that the inner types of the inheritance
hierarchy are free from global constraints'', to ``promotion
refinement is compositional provided global constraints affect only
the inner types of classes with no ascendants''.

To use this relaxation, we need to move the constraint in
\melem{RBQueue} to the superclass, \melem{BQueue}:
\begin{schema}{\SBQueue}[Item]
 \classGen[\oset BQueueCl, BQueue][sBQueue /os, stBQueue
  /oSt]
\where \forall o : sBQueue @ \#(stBQueue~o).items < 2
\end{schema}

\vspace{-9mm}

\begin{zed}
\SRBQueue[Item] ==  \classGen[\oset RBQueueCl,
RBQueue][sRBQueue /os, stRBQueue
  /oSt]
\end{zed}
Now there is no divergence: a superclass extension includes
all objects of its subclasses; all subclass objects are equally
affected by the constraint.  If the precondition on a superclass
object fails, it also fails on the objects of its subclasses.

\vspace{-4mm}
\section{Discussion}
\vspace{-2mm}

\textbf{Behavioural inheritance relaxations.}
This paper showed how over-restrictive traditional refinement constraints can be
 to inheritance: many intuitive
specialisations are not behavioural inheritance refinements in the strict sense. It
highlights the importance of relaxations to the refinement rules:
without them it is very difficult (if not impossible) to
reconcile behavioural inheritance with the flexible scheme
of incremental definition that makes the OO paradigm and OO
inheritance so popular. 

This paper proposes three relaxations to facilitate behavioural inheritance conformance:
\begin{itemize} 
\item The first allows the addition of extra operations to the subclass freely. In~\cite{PhDThesis}, it is 
proved that for any subclass-extra operation
it is possible to find a virtual operation that simulates it and
satisfies the refinement constraints.  This paper argues that there is no harm
in introducing such operations because they are never executed; this
is a property of OO systems that we can rely on. This relaxation is
further confirmed by recent relaxations in other data
refinement settings (see below).
\item The OO abstract class relaxation is perhaps more controversial, and
needs to be applied with care to avoid misunderstandings because it introduces a new kind of refinement contract that differs from the classical one. An OO abstract class operation defines a more \emph{liberal contract}, which effectively binds a behaviour, but allows
subclasses to narrow the precondition. This seems odd because it appears
to allow divergent behaviour, but there is no real divergence
because an OO abstract class has no direct instances.  The objects of
an abstract class are the instances of its subclasses only; its
operations are never executed (they are virtual); collectively, the
objects of an OO abstract class are \emph{polymorphic}: they are allowed to have a multitude of
behaviours that can slightly diverge from each other. As the Queue model and
the other models in \cite{PhDThesis} show, with due caution this
relaxation is extremely useful; it is key in enabling OO inheritance designs that are flexible, make use of polymorphism
and preserve semantic behaviour. 
\item The third relaxation is the result of studying  how
global constraints interfere with local properties. Proving
behavioural conformance at the local level is not the end of the
story. The assurance that the local behavioural conformance
property holds in the global system rests on the compositionality of
promotion with respect to refinement when the promotion is
free~\cite{Lupton:90}. This paper's third relaxation, a design guideline, widens the applicability of the freeness result; the paper argues informally its safety. When this relaxation is
not applicable there is not a practical way to demonstrate
behavioural correctness; refinement proofs of global states are very
complicated even in small systems. This global relaxation has a different nature from the local relaxations given above. Whereas the local relaxations lift certain refinement constraints to allow more refinements, this global one relaxes the proof obligations, extending the freeness rule to more situations to allow more refinement definitions without the need for global proofs.
\end{itemize}

The Queues example presented here illustrates how often inheritance hierarchies are incorrectly assumed to be refinements.
The subclassing of an  unbounded queue by bounded one is a kind of inheritance common in the OO literature that is not, however, a behavioural inheritance refinement. In general, a bounded data type does not refine an
unbounded one. This paper shows that it is possible to demonstrate
behavioural conformance for the Queues inheritance model using the
relaxations and without refactoring the hierarchy; we note, however, that this
is not always possible.  In many cases, the best solution would be to
refactor the hierarchy (\cite{PhDThesis} gives some examples). All behavioural inheritance refinement proofs of the Queues example were automatically proved in Z/Eves. Usually, proofs at the level of inner (or local) types are
trivial; most of them are automatically provable in Z/Eves.

\noindent
\textbf{Behavioural inheritance related concepts.}
This work helps to clarify the relation between
various concepts that have distinct designations in the literature, such
as, behavioural subtyping, behavioural inheritance, data refinement,
class refinement and promotion refinement. The original
concept of behavioural subtyping equates to data refinement in the OO setting, where an
arbitrary refinement relation is allowed. Class refinement extends the theory
of data refinement (which applies to ADTs) to classes; in this work
class refinement equates to Z promotion refinement. Behavioural
inheritance is just one specific class refinement because the
refinement relation is fixed (there may be alternative formulations
of this refinement relation).

\noindent
\textbf{ZOO and other OO models.}
ZOO's high-order OO model with a representation of classes as Z promotions is akin to mathematical models of OO programming languages with a formal semantics. Meyer~\cite{Meyer:97} argues that the OO approach is based on
the mathematical theory of abstract data types; he sees classes as
having a type view and a module view, which correspond to ZOO's
class intension and extension. This means that the results presented here are applicable to other OO settings with a formal semantics, especially those based on design-by-contract, such as Eiffel and JML. ZOO's model, however, differs from first-order models, such as Alloy's~\cite{Jackson:2006aa}, where class fields or attributes are represented as relations; this results in models where everything is global and flat.

\noindent
\textbf{Multiple Inheritance.}
ZOO's style presented here support multiple-inheritance. \cite{PhDThesis} gives a queues example with multiple-inheritance. In the setting of multiple-inheritance, a subclass must be a behavioural-inheritance refinement of all its direct superclasses. 

\vspace{-5mm}
\section{Related Work}
\vspace{-2mm}

It has long been observed the mismatch between the constraints of formal refinement and the needs of more practical software development~\cite{Banach:1998fk}. Retrenchment~\cite{Banach:1998fk,Banach:2007fk} is a more liberal approach to formal-refinement that tries to address this problem. This paper uses this liberalisation idea in the context of mainstream OO inheritance: by studying OO inheritance in the context of data refinement, the paper is able to provide relaxations to the refinement constraints that do not violate the key substitutability principle of both refinement and behavioural inheritance. 

In~\cite{Ref-Reach:Ev-B:2005}, Abrial proposes \textsf{keep} operations (or
actions) to overcome the restrictions of
the \textsf{skip} approach. A \textsf{keep} is a
non-deterministic operation that is guaranteed to preserve the
invariant. Abrial argues that it is safe to add \textsf{keep}
operations to abstract types. This is similar to the superclass virtual operations proposed here, which are safe because they are not visible to the environment.

Whilst the OO model of Liskov and Wing~\cite{Subtyping:LW:1994} is
similar to ZOO's (there is a mapping from objects (atoms) to their
state), their approach is based on a earlier method of data
refinement~\cite{Hoare72:DR} that does not consider initialisation and finalisation.  This paper uses data refinement based on
simulation~\cite{DataRef:HHS:1986}, the enduring basis of the
theory, which accounts for object creation (initialisation) and deletion (finalisation); behavioural conformance cannot be guaranteed if these are not
checked as behaviour of subclass and superclass objects may diverge. Liskov
and Wing's rules correspond to the rule for blocking refinement presented here.
Relaxations are not considered.

ZOO's OO inheritance approach presented here improves Hall's~\cite{Z-OO-Hall:1994}. ZOO represents clearly a class modularly as a promoted ADT and introduces an approach to specify polymorphic operations. ZOO borrows Hall's behavioural inheritance refinement function;  Hall proposes some behavioural inheritance proof rules without a formal proof in the data-refinement setting. Relaxations are not considered.

Wehrheim and Fischer~\cite{Beh-Inh:FW:2000, Beh-Inh:W:2000}
investigate behavioural subtyping in the context of concurrency and
the CSP process algebra. They studied how
extra  subclass operations may interfere with the behaviour of the
superclass as observed from the environment, and under which
conditions are safety and liveness properties preserved by the
subclasses. They propose several inheritance refinement relations;
the more liberal they are, the higher the risk of interference.  The
one that is closer to ZOO's relaxation on extra operations is
\emph{weak subtyping}, which says that the subclass should have the
same behaviour as its superclass as long as no extra operations are
called; the extra operations are not considered in the comparison.
The authors also proposed a more restricted relation, \emph{optimal
subtyping}, which does not allow altering the behaviour of the
superclass at all; it is the same as the \textsf{skip} behaviour.
%The authors argue that weak subtyping is appropriate for exclusive
%access to an object. 

%This is the case with sequential systems, the
%intended application domain of ZOO.

Object-Z~\cite{OZSmith00,Refinement-Z-OZ} defines a formal semantics
for inheritance and a notion of class refinement, but a discussion of behavioural inheritance is
generally absent in its books. In~\cite{Refinement-Z-OZ}, behavioural
inheritance and its relation to refinement is discussed, but no proof
obligations are proposed to check its correctness.
%  In
%\cite{OZSmith00}, there is a Object-Z queues model similar to the one presented here, but there is no comment on the fact that the bounded queue is not behaviour inheritance conformant with the unbounded queue.

%A problem similar to ``interference of global constraints on local structures'' (section~\ref{sec:outer-type}) has been studied by
%M\"uller et al~\cite{Muller:Inv-Layered:2006}. \cite{Muller:Inv-Layered:2006} shows that many invariant proof-checks performed by OO technologies
% only hold in the local realm, and can easily be violated in the global realm with more complex object structures.
% The same problem was highlighted here. This paper showed how
% the behavioural inheritance refinement of \textsf{BQueue}
% by \textsf{RBQueue} proved at the intensional level (the local level)
% could easily by broken by an extra invariant specified at the extensional level (the global level).
% To avoid such problems this paper proposes a design rule that relies on the properties of promotion refinement in Z.

\vspace{-5mm}
\section{Conclusions}
\vspace{-2mm}

This paper investigates behavioural inheritance using ZOO, a Z style of object-orientation, and the theory of Z data refinement. It shows how over-restrictive refinement constraints are to inheritance,
and how important it is to relax such constraints in order to
reconcile incremental definition with behavioural conformance. The paper proposes two new relaxations to the refinement rules that do not compromise the principles of data refinement allowing more refinements than the classical setting. The paper also shows how global properties can
interfere with behavioural inheritance conformance that is only proved locally, and proposes a relaxation to the proof obligations at the global level that allows the important property of composition of promotion with respect to refinement to be more widely applicable in a OO setting, allowing more refinement definitions without the need for global proofs. 
This paper's main contributions are these three relaxations addressing behavioural inheritance conformance, which are the result of a careful examination of OO inheritance in the setting of data refinement. To the author's knowledge, these relaxations have not been proposed before. A secondary contribution is the approach to specify inheritance in Z that improves Hall's approach~\cite{Z-OO-Hall:1994}.

\vspace{-0.6cm}
\paragraph*{Acknowledgements.} Many thanks to Susan Stepney and Fiona Polack for their helpful comments, insight and encouragement on this work.
\vspace{-0.6cm}
\enlargethispage{10mm}

\bibliographystyle{eptcsini}
\bibliography{beh-inh-paper}

\appendix

\end{document}